# Interface-Confined Superconductivity with Thickness-Independent Superfluid Stiffness in (Pb,Sn)Te/FeTe Bilayers

Zi-Jie Yan[1,6], Hongtao Rong[1,6], Yiyuan Luo[2,6], Yufei Zhao[1,3], Pu Xiao[1], Cequn Li[4], Zihao Wang[1], Lok-Kan Lai[1], Annie G. Wang[1], Zhiyuan Xi[1], Yanxing Li[2], Xiaoyu Wei[2], Ke Wang[5], Katja C. Nowack[4], Binghai Yan[1,3], Chih-Kang Shih[2*], and Cui-Zu Chang[1,5*]

[1]Department of Physics, The Pennsylvania State University, University Park, PA 16802, USA

[2]Department of Physics, University of Texas at Austin, Austin, TX 78712, USA

[3]Department of Condensed Matter Physics, Weizmann Institute of Science, Rehovot 7610001, Israel

[4]Laboratory of Atomic and Solid-State Physics, Cornell University, Ithaca, NY 14853, USA

[5]Materials Research Institute, The Pennsylvania State University, University Park, PA 16802, USA

[6]These authors contributed equally: Zi-Jie Yan, Hongtao Rong, and Yiyuan Luo

[*]Corresponding authors: shih@physics.utexas.edu (C.-K. S.); cxc955@psu.edu (C.-Z. C.).

**Abstract: Interface-induced superconductivity in FeTe-based bilayers provides a promising route toward topological superconductivity, yet its spatial extent and dependence on the neighboring layer's topology, symmetry, and electronic structure remain unclear. In this work, we employ molecular beam epitaxy to grow $Pb_{1-x}Sn_xTe$/FeTe bilayers and use angle-resolved photoemission spectroscopy to track the evolution of the $Pb_{1-x}Sn_xTe$ layer from a trivial insulator to a topological crystalline insulator hosting multiple Dirac surface states. Electrical transport reveals robust superconductivity with a nearly constant transition temperature of ≈12 K across the entire composition range ($0 \leq x \leq 1$), despite substantial changes in the electronic structure and band topology of $Pb_{1-x}Sn_xTe$. Double-coil mutual-inductance measurements further reveal comparable superfluid stiffness across the**

**topological phase transition and a nearly thickness-independent superfluid stiffness despite large variations in the constituent-layer thicknesses, providing quantitative evidence for interface-confined superconductivity. These results establish that superconductivity in FeTe-based bilayers is largely insensitive to the topology, crystal symmetry, and detailed electronic structure of the neighboring layer, consistent with Te-assisted removal of interstitial Fe near the interface as a possible microscopic mechanism. The coexistence of interface-confined superconductivity and multiple tunable Dirac surface states in $Pb_{1-x}Sn_xTe$/FeTe bilayers provides a versatile platform for exploring topological superconductivity and interactions among multiple Majorana zero modes.**

**Main text:** Over the past two decades, the search for material platforms capable of hosting topological superconductivity (TSC) has become a central theme in condensed-matter physics because of its close connection to Majorana zero modes (MZMs) and fault-tolerant quantum computation[1-6]. One widely explored route toward TSC is to couple a topological material to a conventional *s*-wave superconductor through the superconducting proximity effect[1,7-9]. Early studies focused primarily on time-reversal-invariant topological insulators, in which a single Dirac surface state (SS) typically acquires superconducting order through proximity coupling to an *s*-wave superconductor. Topological crystalline insulators (TCIs) broaden this framework by hosting multiple Dirac SSs protected by crystalline symmetries rather than solely by time-reversal symmetry[10-17]. When a TCI is coupled to an *s*-wave superconductor[18-21], multiple MZMs may, in principle, coexist within a single vortex core, providing opportunities to investigate their interactions and hybridization[22]. Therefore, establishing a bilayer that combines multiple TCI Dirac SSs with robust superconductivity is essential for exploring the physics of multiple MZMs.

Recent experiments have demonstrated superconductivity in a broad range of FeTe-based bilayers incorporating various telluride layers[23-32], including $(Bi,Sb)_2Te_3$ (refs.[23-27]), Cr-doped $(Bi,Sb)_2Te_3$ (ref.[28]), $MnBi_2Te_4$ (ref.[29]), $CrTe_2$ (ref.[30]), $ZrTe_2$ (ref.[31]), and MnTe (ref.[32]). These discoveries have generated much interest because FeTe is typically antiferromagnetic and nonsuperconducting[33]. In contrast, these bilayers exhibit superconductivity with a transition temperature $T_c$ of 10–12 K. Several mechanisms have been proposed, including interfacial charge transfer[34,35], enhanced spin fluctuations[25], and coupling to Dirac SSs (refs.[25,26]). Superconductivity has also been reported in FeTe films grown directly on MgO and CdTe substrates[36,37], with tensile stress or suppression of monoclinic distortion proposed as possible mechanisms. However, a unified microscopic understanding remains elusive. The neighboring telluride layers in previously studied superconducting FeTe-based bilayers[23-32] predominantly exhibit hexagonal in-plane symmetry, suggesting that lattice symmetry or a specific interfacial geometry might be essential. At the same time, the recurring observation of superconductivity in bilayers containing Te-based neighboring layers suggests that the effect may instead originate primarily from modifications to FeTe itself. The recent observation of superconductivity in stoichiometric FeTe films following the removal of interstitial Fe atoms further supports this possibility[38]. This finding also raises the question of whether superconductivity in FeTe-based bilayers extends throughout the constituent layers or remains confined near the interface. Because zero resistance can arise from a continuous superconducting path, resistive transport alone cannot determine the spatial extent of the superconducting condensate. Measurements of superfluid stiffness $\rho_s$ as a function of film thickness provide a complementary approach to probing its spatial extent. Therefore, a square-lattice telluride overlayer with tunable topological phases, combined with independent control over the thickness of the constituent layers, offers an ideal platform for investigating both the influence of the neighboring layer and the spatial extent of superconductivity.

$Pb_{1-x}Sn_xTe$ has been established as a canonical TCI when it is close to the SnTe end and crystallizes in the rock-salt lattice structure (space group $Fm\bar{3}m$; Fig. 1a). As $x$ increases from 0 to 1, $Pb_{1-x}Sn_xTe$ undergoes a topological phase transition driven by band inversion at the $L$ points (Fig. 1b), evolving from a trivial insulator into a TCI hosting multiple mirror-symmetry-protected Dirac SSs (Fig. 1c)[11,12]. Moreover, unlike the hexagonal neighboring layers in previously studied FeTe-based bilayers, $Pb_{1-x}Sn_xTe$(001) has a square in-plane lattice structure with the same fourfold rotational symmetry as FeTe(001), facilitating the growth of highly ordered epitaxial bilayers. Furthermore, although extensive studies have focused on the (111) surface of $Pb_{1-x}Sn_xTe$ films[17,39-41], relatively few studies have investigated the (001) orientation[42], where the TCI SSs exhibit a distinct electronic structure characterized by hybridized Dirac cones near the $\bar{X}$ points. Therefore, $Pb_{1-x}Sn_xTe$/FeTe bilayers provide a controlled platform for investigating both the influence of neighboring-layer topology and symmetry on superconductivity, as well as the spatial extent of the superconducting condensate, while combining robust superconductivity with multiple Dirac SSs.

In this work, we employ molecular beam epitaxy (MBE) to synthesize a series of bilayer heterostructures consisting of an $m$-unit-cell (UC) $Pb_{1-x}Sn_xTe$ layer and an $n$-UC FeTe layer grown on heat-treated $SrTiO_3$(100) substrates [denoted as an ($m$, $n$) bilayer, Fig. 1f]. In situ angle-resolved photoemission spectroscopy (ARPES) measurements reveal an $x$-driven topological phase transition on the $Pb_{1-x}Sn_xTe$(001) surfaces, accompanied by the emergence of a pair of Dirac cones along the $\bar{X}$-$\bar{\Gamma}$-$\bar{X}$ mirror line near $\bar{X}$ points of the first Brillouin zone (BZ) (Fig. 1d). Interface-induced superconductivity emerges for $m \geq 2$ and $n \geq 5$, becomes progressively enhanced with increasing $m$ or $n$, and ultimately saturates with a $T_c$ of ≈12 K for $m \geq 8$ and $n \geq 15$, independent of $x$. These observations suggest that neither the band topology of the neighboring layer nor the lattice-symmetry mismatch between the two layers is required for the emergence of superconductivity. Furthermore, double-coil mutual-inductance (DCMI) measurements reveal nearly thickness-independent $\rho_s$ for $Pb_{1-x}Sn_xTe$/FeTe bilayers, while the stiffness remains essentially unchanged across the topological phase transition. This behavior demonstrates that the superconducting response is confined near the interface rather than extending throughout the full bilayer thickness. Our results establish the coexistence of superconductivity and hybridized Dirac SSs in $Pb_{1-x}Sn_xTe$/FeTe bilayers, providing important insight into the origin of superconductivity in FeTe-based bilayers, and identify a promising material platform for exploring interactions among multiple MZMs.

**Topological phase transition in $Pb_{1-x}Sn_xTe$/FeTe**

We first characterize a series of MBE-grown (24, 25) bilayers with varying $x$. All samples exhibit sharp, streaky reflection high-energy electron diffraction (RHEED) patterns throughout the growth (Fig. S1), indicating highly ordered crystalline structures. X-ray diffraction (XRD) measurements reveal two sets of sharp diffraction peaks corresponding to the FeTe and $Pb_{1-x}Sn_xTe$ layers, respectively (Fig. 1e). With increasing $x$, the $Pb_{1-x}Sn_xTe$ diffraction peaks systematically shift toward higher $2\theta$ values, yielding a lattice constant that varies linearly with $x$ (Fig. 1h). This linear dependence demonstrates precise control over $Pb_{1-x}Sn_xTe$ stoichiometry. The narrow rocking curves, with full widths at half maximum (FWHMs) of ≈0.092° for the FeTe(003) peak and ≈0.096° for the $Pb_{0.8}Sn_{0.2}Te$(004) peak, further confirm the epitaxial nature of both layers (Fig. S2). In addition, cross-sectional scanning transmission electron microscopy (STEM) images of the $Pb_{0.6}Sn_{0.4}Te$/FeTe bilayer resolve highly ordered crystalline structures in both layers (Fig. 1g), consistent with our RHEED and XRD results. An atomically sharp interface between $Pb_{0.6}Sn_{0.4}Te$ and FeTe is observed, with no detectable elemental intermixing or diffusion, as confirmed by the corresponding energy-dispersive X-ray spectrometry (EDS) measurements (Fig. S3). The high crystalline quality and atomically sharp interface preserve the crystal symmetries of both layers, thereby providing an ideal platform for probing the interplay between TCI Dirac SSs and interface-induced superconductivity.

To investigate the evolution of the electronic structures in $Pb_{1-x}Sn_xTe$/FeTe bilayers, we perform in situ ARPES measurements on a series of (24, 25) bilayers with varying $x$. Because the lattice constant changes by only ≈2% over the entire composition range, $0 \leq x \leq 1$ (Fig. 1h), the BZ dimension of SnTe is used for all samples in Fig. 2. Figure 2a-f shows Fermi-surface maps around the $\bar{X}$ point of the $Pb_{1-x}Sn_xTe$(001) surface BZ for different values of $x$. For $x = 0$ and 0.2, a single electron pocket is observed at $\bar{X}$ (Fig. 2a,b). At $x = 0.4$, the spectral weight at the Fermi level, $E_F$, is strongly suppressed (Fig. 2c), indicating that $E_F$ lies within the bulk band gap. With a further increase in $x$, two pronounced features emerge on opposite sides of $\bar{X}$ and become progressively more separated as $x$ increases (Fig. 2d-f), indicating a substantial reconstruction of the electronic structure. This systematic evolution is consistent with a continuous transition from the topologically trivial phase in PbTe (i.e., $x = 0$) to the TCI phase in SnTe (i.e., $x = 1$). To elucidate the origin of this evolution, we compare the measured band dispersions (Fig. 2g-l) with the band structures obtained from first-principles calculations (Fig. 2m-r). For $x \leq 0.6$, the band dispersions are measured along the $\bar{M}$-$\bar{X}$-$\bar{M}$ direction. The $x = 0$ sample (i.e., PbTe) exhibits a direct band gap of ≈0.23 eV, with $E_F$ intersecting the bulk conduction band (BCB) (Fig. 2g). As $x$ increases, the BCB and bulk valence band (BVB) progressively approach each other, resulting in a continuous reduction of the bulk band gap. At $x = 0.4$, only the BVB is visible below $E_F$, whereas the BCB has shifted above $E_F$ (Fig. 2i), consistent with the suppressed spectral weight observed in the corresponding Fermi-surface map (Fig. 2c). At $x = 0.6$, the gap is further reduced, and the BVB maximum reaches $E_F$ (Fig. 2j). The experimentally observed evolution is well reproduced by our first-principles calculations (Figs. 2m-p and S6), which reveal a continuous decrease in the gap size $\Delta_{gap}$ with increasing $x$ and predict gap closure at a critical composition of $x_c \approx 0.72$ (Fig. 1h). This $x_c$ value exceeds that reported for $Pb_{1-x}Sn_xTe$ single crystals[15,16], likely due to interfacial strain induced by epitaxial growth (Supplementary Information).

Beyond $x_c$, our calculations indicate that the BCB and BVB invert at the $\bar{\Lambda}_1$ and $\bar{\Lambda}_2$ points, driving $Pb_{1-x}Sn_xTe$ into the TCI phase (Fig. 2q,r). Consequently, for $x = 0.8$ and 1.0, the ARPES measurements are performed along the $k_x$ direction through the expected Dirac point at $\bar{\Lambda}_1$. Distinct linearly dispersing states emerge within the inverted bulk gap, as marked by the black arrows in Fig. 2k,l. These states agree well with the calculated Dirac SSs (Fig. 2q,r). Our calculations further indicate that the Dirac points lie at ≈0.11 eV above $E_F$ for $x = 0.8$ and ≈0.15 eV above $E_F$ for $x = 1.0$. The emergence of these Dirac SSs, together with the observed evolution of the bulk gap and band inversion, provides direct evidence for a topological phase transition in the $Pb_{1-x}Sn_xTe$/FeTe bilayers. Therefore, our ARPES measurements establish an $x$-driven topological phase transition in these $Pb_{1-x}Sn_xTe$/FeTe bilayers and demonstrate that the TCI Dirac SSs remain well preserved in these bilayers with $x \geq 0.8$.

**Robust superconductivity in $Pb_{1-x}Sn_xTe$/FeTe**

Next, we investigate interface-induced superconductivity in (24, 25) bilayers with varying $x$. Figure 3a shows the temperature $T$-dependent sheet longitudinal resistance $R_{xx}$ of a series of (24, 25) bilayers spanning the entire composition range, $0 \leq x \leq 1$. Remarkably, all bilayers exhibit sharp superconducting transitions with nearly identical $T_c \approx 12.0$ K (Fig. 3d), despite the substantial evolution of the electronic structure revealed by ARPES (Fig. 2). These $T_c$ values are comparable with those reported for other FeTe-based bilayers[23-26,28-32]. More importantly, the persistence of superconductivity across the topological phase transition demonstrates that its emergence is not directly correlated with the topological property of the $Pb_{1-x}Sn_xTe$ layer. Together with the square-lattice symmetry of $Pb_{1-x}Sn_xTe(001)$, this observation suggests that neither band topology nor lattice symmetry is necessary for inducing superconductivity in FeTe-based bilayers. More generally, superconductivity has thus far been observed whenever FeTe is interfaced with a Te-based compound[23-32], despite substantial differences in electronic structure, band topology, magnetism, and crystal symmetry of the neighboring layers. This empirical trend points to a common origin, primarily associated with the FeTe layer, a possibility we investigate further below.

To further examine the influence of $x$ on the superconducting state, we perform DCMI measurements[43,44], a well-established technique for probing the macroscopic superfluid properties in a wide range of superconductors[45-53]. Briefly, the DCMI measurement employs a pair of concentric coils positioned close to the sample surface (Figs. 3b and S4a). The drive coil generates an AC magnetic field, while the pickup coil detects the magnetic screening response generated by Meissner supercurrents in the sample. This response is directly related to the sample's complex sheet conductivity. It can be used to extract key superconducting parameters, including the superfluid density $n_s$ and the superfluid stiffness $\rho_s$. Because the effective superconducting thickness of the $Pb_{1-x}Sn_xTe$/FeTe bilayers is not known a priori, we focus on $\rho_s$ rather than $n_s$. Whereas $n_s$ depends explicitly on the assumed superconducting volume, $\rho_s$ directly quantifies the integrated response of the entire superconducting region within the sample and can therefore be determined without introducing uncertainties associated with the unknown exact superconducting thickness (Methods).

Figure 3c shows the $T$ dependence of $\rho_s$ for the (24, 25) bilayers with different $x$. In all samples, $\rho_s$ remains negligible for $T \geq T_c$ and exhibits a pronounced upturn between $T = 10$ K and $T = 11.5$ K, indicating the establishment of macroscopic phase coherence throughout the superconducting region. The onset temperatures of the $\rho_s$-$T$ curves are close to, but consistently slightly lower than, the $T_c$ values determined from our electrical transport measurements (Fig. 3a). Such a discrepancy is expected and has been reported in prior studies[47,54]. Transport measurements identify $T_c$ by forming a continuous superconducting path between the source and drain electrodes, whereas DCMI measurements probe the global diamagnetic screening response of the entire sample. Although the $\rho_s$-$T$ curves do not quantitatively overlap, they exhibit remarkably similar $T$ dependence, consistent with prior studies on FeSe (refs.[50,51]), and reach comparable values of $\rho_s$ at low temperatures across the entire composition range, $0 \leq x \leq 1$ (Fig. 3d). The absence of any systematic evolution in either the shape or magnitude of the $\rho_s$-$T$ curves indicates that the superfluid stiffness in $Pb_{1-x}Sn_xTe$/FeTe is largely independent of $x$, in agreement with the electrical transport measurements. In particular, no monotonic variation associated with the topological phase transition of the $Pb_{1-x}Sn_xTe$ layer is observed. The moderate sample-to-sample variations in the absolute magnitude of $\rho_s$ show no systematic dependence on $x$ and are therefore unlikely to reflect an intrinsic electronic trend.

**Thickness-independent superfluid stiffness in $Pb_{1-x}Sn_xTe$/FeTe**

To further investigate the origin of the emergent superconductivity in $Pb_{1-x}Sn_xTe$/FeTe bilayers, we examine its thickness dependence in two series of ($m$,$n$) bilayers with varying $m$ and $n$. Figure 4a shows the $R_{xx}$-$T$ curves of the ($m$, 25) bilayers. For $m = 0$, corresponding to a bare 25-UC FeTe film, a resistance hump appears near $T \approx 59$ K, associated with the paramagnetic-to-antiferromagnetic phase transition of FeTe and consistent with prior studies[23,25,30]. Upon deposition of 2-UC SnTe, superconductivity emerges, with $T_{c,\text{onset}} = 10.8$ K and $T_{c,0} = 2.9$ K. As $m$ increases, both $T_{c,\text{onset}}$ and $T_{c,0}$ increase, accompanied by a monotonic reduction in their difference (i.e., $\Delta T_c = T_{c,\text{onset}} - T_{c,0}$), suggesting a progressively sharper superconducting transition with increasing $m$.

However, for $m \geq 8$, further increases in $m$ produce little change in the superconducting properties, with $T_c$ saturating at ≈11.7 K (Fig. 4c). Similar behavior is observed in the (24, $n$) bilayers (Fig. 4b). The (24, 5) bilayer exhibits a superconducting transition with $T_{c,onset}$ = 11.1 K and $T_{c,0}$ = 3.4 K. Increasing $n$ sharpens the superconducting transition and raises $T_{c,0}$. In comparison, $T_c$ gradually saturates at ≈11.9 K for $n \geq 15$ (Fig. 4c). Similar thickness-dependent behavior has been reported in other FeTe-based bilayers incorporating a variety of neighboring Te-based compounds[28-30]. The nearly identical evolution and saturation of $T_c$, despite substantial differences in the electronic, magnetic, topological, and crystalline properties of the neighboring layers, point to a common superconducting mechanism in all these FeTe-based bilayers.

To further probe the spatial extent of superconductivity, we perform DCMI measurements on three SnTe/FeTe bilayers with substantially different $m$ and $n$ but comparable $T_c$ values. Because $\rho_s$ measures the integrated phase-coherent superfluid response of the entire superconducting region within the sample, it would be expected to increase with the total bilayer thickness ($m$ + $n$) if superconductivity is uniformly distributed throughout the bilayer. Remarkably, however, the $\rho_s$-$T$ curves nearly overlap over the entire measured $T$ range despite large variations in both $m$ and $n$ (Fig. 4d). This result indicates that increasing either $m$ or $n$ contributes negligibly to the overall superconducting response. These observations provide compelling evidence that superconductivity is confined to a limited vertical region rather than extending throughout the full thickness of the SnTe/FeTe bilayer. Given that superconductivity emerges only when SnTe and FeTe are brought into contact, the most natural interpretation is that the superconducting state is confined near the SnTe/FeTe interface.

To estimate the effective superconducting thickness, we compare the superfluid stiffness $\rho_s$ of the $Pb_{1-x}Sn_xTe$/FeTe bilayers with that of 20-UC and 40-UC stoichiometric FeTe films[38,55], measured using scanning superconducting quantum interference device (SQUID) microscopy (Fig. S8), a technique complementary to DCMI. In contrast to the nearly thickness-independent $\rho_s$ observed in the $Pb_{1-x}Sn_xTe$/FeTe bilayers, the 40-UC FeTe film exhibits approximately twice the $\rho_s$ of the 20-UC film at locations with comparable local $T_c$ (Fig. S8b), consistent with an approximately linear dependence of $\rho_s$ on FeTe thickness over this range. Assuming comparable $\rho_s$ per superconducting FeTe UC in both systems, comparison with these stoichiometric FeTe films yields an effective superconducting thickness of ≈4–8 FeTe UC (≈2.5–5 nm) for the SnTe/FeTe bilayers (Fig. 4d), further supporting predominantly interface-confined superconductivity. The assumptions and limitations of this estimate are discussed in the Supplementary Information.

**Discussion and outlook**

Our observations provide quantitative evidence for predominantly interface-confined superconductivity in $Pb_{1-x}Sn_xTe$/FeTe bilayers. A possible microscopic explanation is suggested by our recent study of stoichiometric FeTe films[38], which demonstrated that superconductivity could emerge in FeTe after excess interstitial Fe atoms are removed through post-growth Te annealing. During the MBE growth of the top $Pb_{1-x}Sn_xTe$ layer, Pb, Sn, and Te are co-evaporated onto the bottom FeTe layer, with the Te flux typically exceeding the Pb and Sn fluxes by a factor of 5-10. Consequently, the FeTe surface is inevitably exposed to a substantial excess Te flux before it is fully covered by the $Pb_{1-x}Sn_xTe$ layer. This exposure may serve as an effective in situ Te-annealing process, reducing excess interstitial Fe atoms near the FeTe surface and thereby promoting superconductivity. Unlike intentional post-growth Te annealing in our recent study[38], which can modify the entire FeTe film, this unintentional annealing process is expected to occur primarily during the initial stages of $Pb_{1-x}Sn_xTe$ growth, before the FeTe surface becomes fully covered. Therefore, only a limited vertical region near the $Pb_{1-x}Sn_xTe$/FeTe interface may undergo sufficient excess Fe removal to become superconducting. This picture naturally accounts for the interface-confined superconductivity inferred from our experimental observations, including the robustness of superconductivity against variations in topology and lattice symmetry (Fig. 3), the saturation of $T_c$ with increasing $m$ (Fig. 4c), and the nearly thickness-independent $\rho_s$ (Fig. 4d). More broadly, it provides a unified framework for understanding why superconductivity emerges in FeTe-based bilayers whose neighboring layers possess widely different electronic structures, topological properties, magnetic orders, and crystal symmetries. Spatial variations in the initial interstitial Fe concentration or the efficiency of its removal may also contribute to the superconducting inhomogeneity frequently observed in FeTe-based bilayers[30].

To summarize, we use MBE to grow a series of $Pb_{1-x}Sn_xTe$/FeTe bilayers with varying $x$ and systematically investigate their electronic structure and superconducting properties using ARPES, electrical transport, and DCMI measurements, complemented by scanning SQUID measurements on stoichiometric FeTe films. ARPES reveals an $x$-driven topological phase transition in the $Pb_{1-x}Sn_xTe$ layer, accompanied by the emergence of multiple Dirac SSs, in good agreement with our first-principles calculations. Despite the substantial evolution of the electronic structure across the topological phase transition, all $Pb_{1-x}Sn_xTe$/FeTe bilayers with varying $x$ exhibit nearly identical $T_c$ values and comparable $\rho_s$, demonstrating that superconductivity in FeTe-based bilayers is largely insensitive to the band topology, crystal symmetry, and detailed electronic structure of the neighboring layer. Furthermore, the nearly thickness-independent $\rho_s$ observed in SnTe/FeTe bilayers with substantially different $m$ or $n$ provides strong evidence that the superconducting state is confined to the SnTe/FeTe interface. Together with our recent studies of stoichiometric FeTe films[38,55], these observations suggest a common microscopic mechanism involving modifications of FeTe near the interface, potentially driven by Te-assisted removal of excess interstitial Fe atoms during bilayer growth. Finally, the coexistence of interface-confined superconductivity and multiple Dirac SSs in $Pb_{1-x}Sn_xTe$/FeTe bilayers establishes a promising platform for exploring TSC and the possible interactions and hybridization among multiple MZMs.

## Methods

### MBE growth

All $Pb_{1-x}Sn_xTe$/FeTe bilayers studied in this work are grown in a commercial MBE chamber (Lab 10, ScientaOmicron) with a base vacuum better than $3 \times 10^{-10}$ mbar. Both metallic 0.5% Nb-doped $SrTiO_3$(100) and insulating $SrTiO_3$(100) substrates are used. FeTe films grown on metallic 0.5% Nb-doped $SrTiO_3$(100) are used for ARPES measurements, while those grown on insulating $SrTiO_3$(100) are used for ex situ XRD, STEM, electrical transport, and DCMI measurements. Before MBE growth, all $SrTiO_3$(100) substrates are soaked in hot deionized water (≈80 °C) for 2 hours, immersed in a ≈4.5% HCl solution for 2 hours, and annealed at ≈974 °C for 3 hours in a tube furnace with flowing oxygen. These treatments passivate and reconstruct the $SrTiO_3$(100) surface, making it suitable for FeTe growth. The heat-treated $SrTiO_3$(100) substrates are subsequently loaded into the MBE chamber and outgassed at ≈600 °C for 1 hour. High-purity Fe (99.995%), Te (99.9999%), Pb (99.999%), and Sn (99.999%) are co-evaporated from Knudsen effusion cells. The growth temperature is ≈330 °C for FeTe and ≈280 °C for $Pb_{1-x}Sn_xTe$. The growth process is monitored by RHEED (Fig. S1). The growth rates are ≈0.16 UC per minute for FeTe and ≈0.4 UC per minute for $Pb_{1-x}Sn_xTe$, as calibrated by ex situ atomic force microscopy (AFM) and STEM measurements. No capping layer is used in any measurement.

### ARPES measurements

In situ ARPES measurements are performed in a chamber with a base vacuum better than $5 \times 10^{-11}$

mbar. After MBE growth, the $Pb_{1-x}Sn_xTe$/FeTe bilayers are transferred directly from the MBE chamber to the ARPES chamber without breaking the ultrahigh vacuum. A DA30L electron analyzer (ScientaOmicron) is used with a helium discharge lamp (He $I_\alpha$, $h\nu$ = 21.218 eV). The energy and angle resolutions are set to ≈10 meV and ≈0.1°, respectively. ARPES measurements on samples with $x$ = 0.0, 0.2, and 0.4 (Fig. 2g-i) are performed at $T$ = 30 K. To access the band dispersion above the Fermi level, ARPES measurements on samples with $x$ = 0.6, 0.8, and 1.0 (Fig. 2j-l) are performed at $T$ = 300 K.

**Electrical transport measurements**

The $Pb_{1-x}Sn_xTe$/FeTe bilayers grown on 3 mm × 10 mm heat-treated $SrTiO_3$(100) substrates are patterned into Hall-bar geometries by scratching using a computer-controlled motorized probe station. The effective Hall-bar dimensions are ≈1 mm × 0.5 mm. Electrical contacts are made by pressing indium spheres on the Hall bar. Electrical transport measurements are performed in a Physical Property Measurement System (PPMS; Quantum Design DynaCool, 1.7 K, 9 T). An excitation current of 1 μA is used in all $R_{xx}$-$T$ measurements. To minimize oxidation, all bilayer samples are measured within 30 minutes after removal from the MBE chamber. For each sample, the superconducting onset temperature $T_{c,onset}$ is defined as the temperature at which $R_{xx}$ begins to decrease rapidly. The superconducting transition temperature $T_c$ is defined as the temperature at which $R_{xx}$ decreases to 50% of its normal-state resistance at $T = T_{c,onset}$, whereas the zero-resistance temperature $T_{c,0}$ is defined as the temperature at which $R_{xx}$ decreases to 1% of its normal-state resistance at $T = T_{c,onset}$. The error bars in $T_c$ shown in Figs. 3d and 4c are defined as the difference between $T_{c,onset}$ and $T_{c,0}$, i.e., $\Delta T_c = T_{c,onset} - T_{c,0}$.

**DCMI measurements**

The superfluid density and superfluid stiffness of the $Pb_{1-x}Sn_xTe$/FeTe bilayers are measured using a DCMI setup[44], consisting of a concentric double coil probe comprising a dipole drive coil and a quadrupole pickup coil wound on a common ceramic rod (Fig. 3b). The probe is mounted on a Pan-type piezo walker, and the whole apparatus is attached to a cold head connected to a liquid-helium dewar set inside an ultrahigh vacuum chamber with a base vacuum of $3 \times 10^{-10}$ mbar. When an alternating current (AC) is applied to the drive coil, a screening supercurrent is induced in the bilayer sample through the Meissner effect. The resulting magnetic response is detected by the pickup coil and is determined by the complex sheet conductivity $G = (\sigma_1 + i\sigma_2)d$, where $d$ is the effective superconducting thickness[43-46,48,49]. During the DCMI measurements, the coil-to-sample distance $h$ is kept at 60–70 μm (Fig. S4). The probe has an outer diameter of ≈1.5 mm, whereas the effective area of each sample is ≈3 mm × 8 mm, ensuring that the measured response originates entirely from the $Pb_{1-x}Sn_xTe$/FeTe bilayer and reflects its macroscopic superconducting properties. Before each DCMI measurement, the $Pb_{1-x}Sn_xTe$/FeTe bilayer is held at $T = 4.3$ K for 3 hours to minimize thermal drift. An SR830 lock-in amplifier is used to apply an AC drive current $I_D$ of ≈1.5 μA to the drive coil and simultaneously measure the complex pickup coil voltage $\delta V$ (Fig. S5). The relationship between $\delta V$ and $G$ is given by[43,44,46,49]:

$$\delta V = i\omega I_D \int_0^\infty dx \frac{M(x)}{1 - \frac{2}{i\omega\mu_0 hG}x}, \tag{1}$$

where $\omega$ is the driving angular frequency, and $\mu_0$ is the vacuum permeability. The function $M(x)$ is defined as $M(x) = \pi\mu_0 h\xi(x)$, where $\xi(x)$ is a dimensionless geometric factor determined by the coil dimensions[44]. To extract $G$ from $\delta V$, we construct a complex grid of $G(\sigma_1, i\sigma_2)$ and calculate the corresponding $\delta V$ values, [Re($\delta V$), Im($\delta V$)], over the grid[44]. The experimentally measured $\delta V$ is then mapped onto this grid using a numerical script to obtain the corresponding $G$. Within the thin-film limit, $\lambda_L \gg d$, assuming a homogeneous superconductor with spatially uniform effective mass $m^*$ and superfluid density $n_s$, the imaginary part of $G$ is directly related to $n_s$, London penetration depth $\lambda_L$, and superfluid stiffness $\rho_s$ through[43-46,48,49]:

$$n_s = \frac{m^* \omega \mathrm{Im}(G)}{e^2 d} \tag{2}$$

$$\frac{1}{\lambda_L^2} = \frac{\mu_0 \omega \mathrm{Im}(G)}{d} \propto n_s \tag{3}$$

$$\rho_s = \frac{\hbar^2 \omega \mathrm{Im}(G)}{4 k_B e^2} \propto \frac{n_s d}{m^*} \tag{4}$$

where $\hbar$ is the reduced Planck constant and $k_B$ is the Boltzmann constant. More generally, for a superconducting bilayer with spatially varying $n_s(z)$ and $m^*(z)$, the imaginary part of $G$ can be expressed as an integral of the local superfluid response along the out-of-plane direction:

$$\omega \mathrm{Im}(G) = e^2 \int dz \frac{n_s(z)}{m^*(z)}. \tag{5}$$

The corresponding superfluid stiffness $\rho_s$ is then given by

$$\rho_s = \frac{\hbar^2 \omega \mathrm{Im}(G)}{4 k_B e^2} = \frac{\hbar^2}{4 k_B} \int_0^d dz \frac{n_s(z)}{m^*(z)}. \tag{6}$$

Unlike $n_s$, which depends explicitly on the assumed superconducting volume through $d$, $\rho_s$ reflects the integrated phase-coherent superconducting response of the sample and can be determined without prior knowledge of the effective superconducting thickness. Because superconductivity in $Pb_{1-x}Sn_xTe$/FeTe bilayers may be confined near the interface and may involve multiple superconducting channels, $\rho_s$ provides a more reliable metric for quantitatively comparing superconducting properties among different $Pb_{1-x}Sn_xTe$/FeTe bilayers. For the thickness-dependent analysis presented in the main text, we assume that the average $n_s(z)/m^*(z)$ within the superconducting region is comparable across SnTe/FeTe bilayers with different thicknesses. Under this assumption, $\rho_s$ scales approximately linearly with $d$ at a given temperature. The justification and limitations of this assumption are discussed in the Supplementary Information.

**Scanning SQUID measurements**

Scanning SQUID measurements are performed using a SQUID susceptometer comprising a pickup loop (inner/outer radii of 0.4/1 μm) coupled to a SQUID circuit and a concentric field coil (inner/outer radii of 1.5/3 μm). The FeTe samples are mounted on a temperature-controlled stage thermally anchored to the mixing chamber of a Bluefors dilution refrigerator (LD400), with a Cernox thermometer monitoring the sample temperature. During the measurements, the SQUID–sample separation is maintained at ≈1 μm. We apply an AC of 0.5 mA at ≈213.7 Hz to the field coil, generating a local magnetic field of ≈50 μT and inducing Meissner screening currents in the FeTe film. The resulting screening flux is detected by the pickup loop through a change in the mutual inductance, $\Delta M$, between the field coil and pickup loop. In the Pearl limit ($d \ll \lambda_L$), $\Delta M$ is proportional to $d/\lambda_L^2$ at fixed SQUID–sample distance and thus directly probes the sheet superfluid stiffness. The London penetration depth $\lambda_L$ is extracted by fitting the measured mutual-inductance response to a thin-film screening model[55,56]. The corresponding superfluid stiffness $\rho_s$ is then calculated as

$$\rho_s = \frac{\hbar^2 d}{4\mu_0 k_B e^2 \lambda_L^2}. \quad (7)$$

**XRD measurements**

High-resolution XRD measurements are performed at room temperature using a Malvern Panalytical X'Pert[3] materials research diffractometer (MRD) with Cu-$K_{\alpha 1}$ radiation (wavelength $\lambda \approx 1.5405980$ Å) in the $\theta$-$2\theta$ geometry. The diffraction peaks are indexed according to the tetragonal crystal structure of FeTe (space group $P4/nmm$) and the rock-salt structure of $Pb_{1-x}Sn_xTe$ (space group $Fm\bar{3}m$), confirming the high crystallinity and phase uniformity of the MBE-grown $Pb_{1-x}Sn_xTe$/FeTe bilayers.

**STEM measurements**

The $Pb_{1-x}Sn_xTe$/FeTe bilayers used in our cross-sectional STEM measurements are prepared using FEI Helios 660 Nanolab and FEI Scios 2 focused ion beam (FIB) systems. To minimize damage during FIB processing, we deposit a 400-nm-thick amorphous carbon layer onto the sample surface using a Leica EM ACE200 sputter coater. Aberration-corrected high-angle annular dark-field scanning transmission electron microscopy (HAADF-STEM) measurements are performed using an FEI Titan$^3$ G2 S/TEM operated at an accelerating voltage of 300 kV. All ADF-STEM images are acquired with a probe convergence angle of ≈30 mrad, a probe current of ≈70 pA, and collection angles of 52–253 mrad.

**First-principles calculations**

The first-principles calculations are performed using the Vienna Ab initio Simulation Package (VASP) within the projector-augmented wave method[57]. Exchange-correlation effects are treated using the generalized gradient approximation (GGA) in the Perdew-Burke-Ernzerhof (PBE) form[58], as well as the Heyd-Scuseria-Ernzerhof (HSE) screened hybrid functionals HSE03 (ref.[59]) and HSE06 (ref.[60]) for comparison. Alloying is modeled using the virtual crystal approximation (VCA) (ref.[61]) together with the lattice constants obtained from our XRD measurements. BZ integration is performed using a 15×15×15 Γ-centered *k*-point mesh. Maximally localized Wannier functions for the Sn-*p*, Te-*p* orbitals are constructed using the Wannier90 package[62]. The Dirac SSs are calculated using the recursive Green's function method for a semi-infinite geometry without explicitly including the bottom FeTe layer. In our ARPES measurements, the $Pb_{1-x}Sn_xTe$ layer is ≈15 nm thick, substantially exceeding the inelastic mean free path of the photoelectrons ($\leqslant$0.6 nm) (ref.[63]). Consequently, the measured ARPES signal predominantly probes the near-surface low-energy electronic states of $Pb_{1-x}Sn_xTe$, with negligible contribution from the buried FeTe layer. Therefore, we compare the experimental ARPES dispersions (Fig. 2g-l) with the calculated surface-projected electronic structures of $Pb_{1-x}Sn_xTe$ (Fig. 2m-r), which reproduce the principal features observed experimentally.

The $Pb_{1-x}Sn_xTe$(001) surface belongs to the $C_{4v}$ point group, in which four mirror planes intersect at $\bar{\Gamma}$ and protect the surface Dirac cones (Fig. 1b,c). Our VCA-based calculations reveal a topological phase transition from trivial PbTe (i.e., $x = 0$) to nontrivial SnTe (i.e., $x = 1$). We also calculate the band gap at the $L$ point in $Pb_{1-x}Sn_xTe$ as a function of $x$ (Figs. 1h and S6). Different exchange-correlation functionals yield different critical Sn concentrations $x_c$, defined by $\Delta_{gap}(x_c) = 0$. Within the PBE functional, $x_c$ is ≈0.2, consistent with a prior study[14]. Compared with the experimental band gaps extracted from ARPES measurements on PbTe and $Pb_{0.8}Sn_{0.2}Te$ (Figs. 2g,h and S6), PBE underestimates the gap, whereas HSE06 overestimates it. HSE03 slightly overestimates the gap but best agrees with our experiments overall. Because Wannier90 cannot be applied directly within the VCA framework, we adjust the spin-orbit coupling (SOC) strength in subsequent PBE calculations to reproduce the HSE03 band gaps and then construct Wannier-based tight-binding models. The PbTe pseudopotential is used to simulate $Pb_{1-x}Sn_xTe$, with the SOC strength set to 0.83 ($x = 0$), 0.86 ($x = 0.2$), 0.89 ($x = 0.4$), 0.91 ($x = 0.6$), 0.95 ($x = 0.8$), and 0.99 ($x = 1$), respectively. The lattice constants are taken from our XRD measurements (Fig. 1h). We find that the bulk gap gradually closes and reopens, signaling the topological phase transition. At the same time, Dirac SSs emerge within the inverted gap (Fig. 2m-r). The gapless surface Dirac point appears along the $\bar{\Gamma}$-$\bar{X}$-$\bar{\Gamma}$ direction instead of the $\bar{M}$-$\bar{X}$-$\bar{M}$ direction because $\bar{\Gamma}$-$\bar{X}$-$\bar{\Gamma}$ lies within a mirror plane. Figure S7f further shows that two surface Dirac cones coexist near the $\bar{X}$ point in SnTe as a consequence of time-reversal symmetry, consistent with the calculated mirror Chern number $n_{\mathrm{M}} = 2$. Along the $\bar{\mathrm{M}}'$-$\bar{\Lambda}_1$-$\bar{\mathrm{M}}'$ direction measured by ARPES (Fig. 2l), a single Dirac cone located on the mirror plane is observed.

**Acknowledgments:** We thank C. X. Liu, A. H. MacDonald, H. X. Tan, W. D. Wu, H. M. Yi, and J. Zhu for helpful discussions.

**Funding:** This work is primarily supported by the ONR award (N000142412133), including the MBE growth and PPMS measurements. The AFM and XRD measurements are partially supported by the DOE grant (DE-SC0023113). The STEM measurements are partially supported by the Penn State MRSEC for Nanoscale Science (DMR-2011839) and the Gordon and Betty Moore Foundation (GBMF14458, DOI: 10.37807/GBMF14458). The DCMI measurements performed at UT Austin are partially supported by the NSF grant (DMR-2219610), the Welch Foundation grant (F-2164), and the AFOSR award (FA2386-21-1-4061). The scanning SQUID measurements conducted at Cornell University are supported by the AFOSR award (FA9550-21-1-0429). C.-Z. C. acknowledges the support from the Gordon and Betty Moore Foundation's EPiQS Initiative (Grant GBMF9063 to C.-Z. C.).

**Author contributions:** CZC and CKS conceived and designed the experiments. ZJY, HR, PX, ZW, LKL, and CZC performed the MBE growth. ZJY, HR, PX, ZW, LKL, AGW, ZX, and CZC performed the electrical transport measurements. HR, ZJY, and CZC performed the ARPES measurements. ZJY and CZC performed the XRD measurements. YLuo, YLi, XW, and CKS performed the DCMI measurements. KW, ZJY, and HR performed the STEM measurements. CL and KCN performed the scanning SQUID measurements. YZ and BY provided theoretical support. ZJY, HR, YLuo, CKS, and CZC analyzed the data and wrote the manuscript with input from all authors.

**Competing interests:** The authors declare no competing interests.

**Data availability:** The data that support the findings of this study are openly available at Zenodo (https://doi.org/10.5281/zenodo.21150459) and are also available from the corresponding authors upon request.

**Figures and figure captions:**

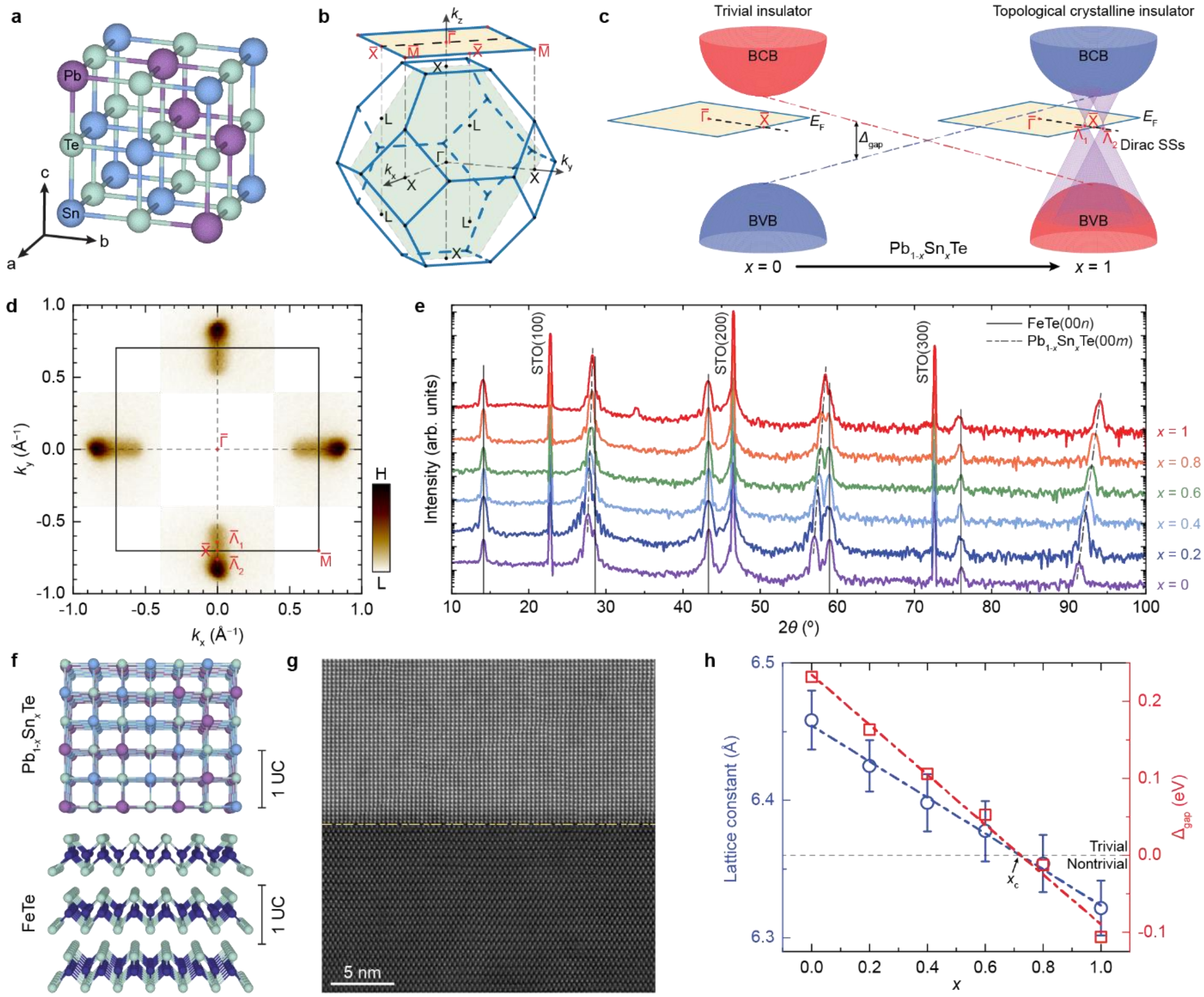

**Fig. 1| MBE-grown $Pb_{1-x}Sn_xTe$/FeTe bilayers.** **a**, Crystal structure of $Pb_{1-x}Sn_xTe$. **b**, First BZ of $Pb_{1-x}Sn_xTe$ in 3D and its 2D projection onto the (001) surface (yellow). The BZ is invariant under reflection about the (110) mirror plane (light green), which projects onto the (001) surface BZ as the $\overline{X}$-$\overline{\Gamma}$-$\overline{X}$ mirror line (black dashed line). **c**, Schematic evolution of the electronic band structure of $Pb_{1-x}Sn_xTe$. As $x$ increases from 0 to 1, band inversion drives $Pb_{1-x}Sn_xTe$ from a trivial insulator into a TCI phase hosting two Dirac SSs located at the $\overline{\Lambda}_1$ and $\overline{\Lambda}_2$ points along the $\overline{X}$-$\overline{\Gamma}$-$\overline{X}$ mirror line

near $\overline{\mathrm{X}}$. **d**, Fermi surface map of a 24-UC SnTe/25-UC FeTe bilayer. The black square marks the first BZ of SnTe. The Fermi surface map is obtained by integrating the spectral weight within an energy window of [-50, 50] meV near $E_F$. **e**, XRD spectra of the (24, 25) bilayers with different $x$. **f**, Schematic of the $Pb_{1-x}Sn_xTe$/FeTe bilayer. **g**, Cross-sectional STEM image of the (24, 25) bilayer with $x = 0.4$. **h**, $x$ dependence of the lattice constant (blue) and band gap size $\Delta_{gap}$ (red) of $Pb_{1-x}Sn_xTe$. As $x$ increases from 0 to 1, $Pb_{1-x}Sn_xTe$ undergoes a topological phase transition at $x_c = 0.72$, where $\Delta_{gap} = 0$. The lattice constants are extracted by fitting the $Pb_{1-x}Sn_xTe$(006) XRD peaks in (**e**), and the $\Delta_{gap}$ values are acquired by our theoretical calculations.

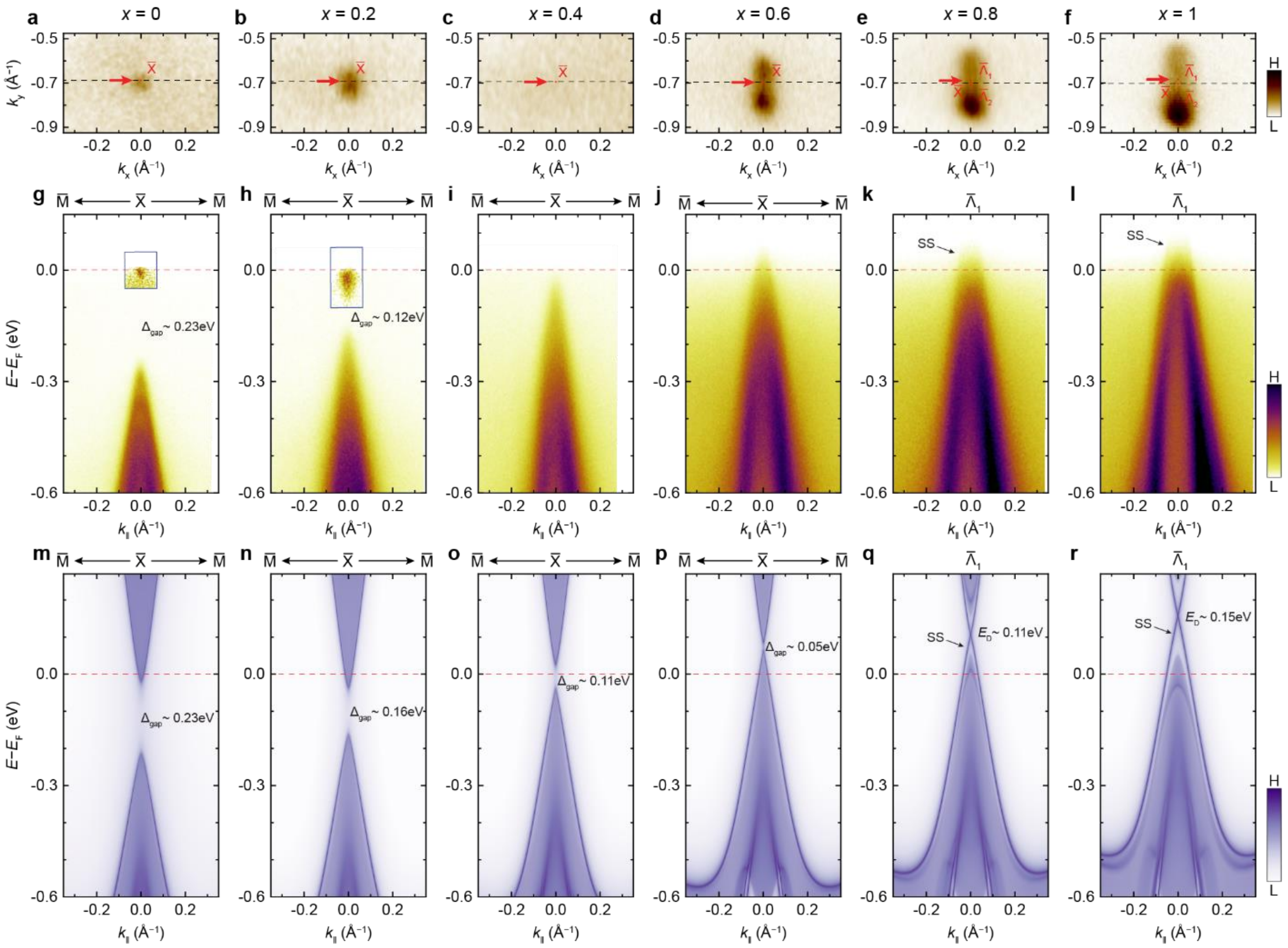

**Fig. 2| Evolution of the electronic band structure of the (24, 25) bilayers with increasing *x*. a-f**, Fermi surface maps of $Pb_{1-x}Sn_xTe$/FeTe bilayers near the $\bar{X}$ point. All Fermi surface maps are obtained by integrating the spectral weight within an energy window of [-50, 50] meV near $E_F$. The red arrows in (**a-f**) indicate the momentum directions along which the band dispersions are measured in (**g-l**) and calculated in (**m-r**), respectively. **g-j**, Band dispersions of $Pb_{1-x}Sn_xTe$/FeTe bilayers with $0 \leq x \leq 0.6$ along the $\bar{M}$-$\bar{X}$-$\bar{M}$ direction. **k**,**l**, Band dispersions of $Pb_{1-x}Sn_xTe$/FeTe bilayers with $0.8 \leq x \leq 1$ around the $\bar{\Lambda}_1$ point along the $k_x$ direction. **m-p**, Calculated band dispersions of $Pb_{1-x}Sn_xTe$(001) surfaces with $0 \leq x \leq 0.6$ along the $\bar{M}$-$\bar{X}$-$\bar{M}$ direction. **q**,**r**, Calculated band dispersions of $Pb_{1-x}Sn_xTe$(001) surfaces with $0.8 \leq x \leq 1$ around the $\bar{\Lambda}_1$ point along the $k_x$ direction. ARPES measurements on samples with $x$ = 0.0, 0.2, and 0.4 (**a-c** and **g-i**) are performed at $T$ = 30 K, whereas those on samples with $x$ = 0.6, 0.8, and 1.0 (**d-f** and **j-l**) are performed at $T$ = 300 K. The calculations in (**m-r**) do not include the bottom FeTe layer.

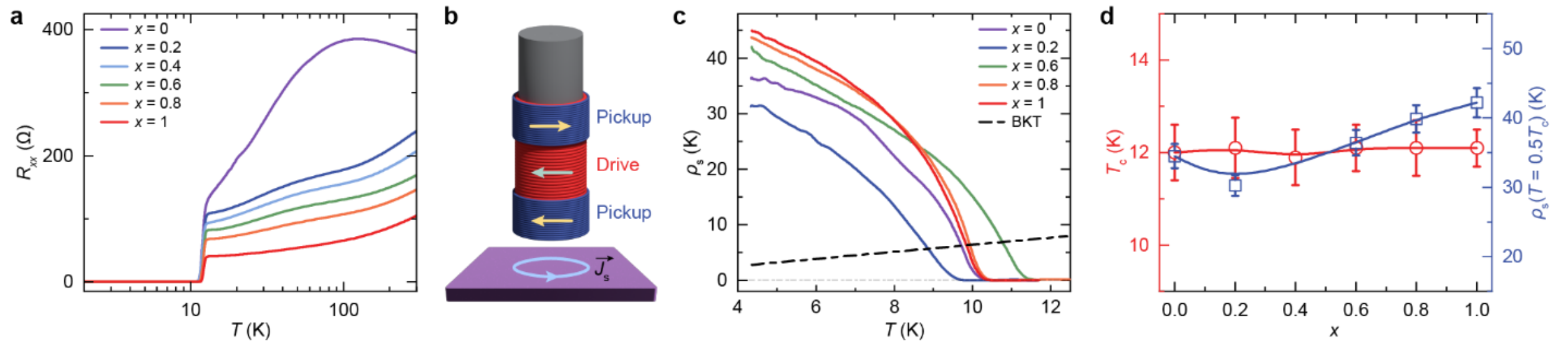

**Fig. 3| Robust superconductivity in the (24, 25) bilayers with varying *x*.** **a**, $R_{xx}$-$T$ curves. **b**, Schematic of the DCMI setup. A dipole inner coil (drive coil, red) and a quadrupole outer coil (pickup coil, blue) are wound concentrically onto a ceramic rod (gray). This setup is positioned close to the sample surface (purple), and the arrows indicate the coil winding directions. **c**, $T$ dependence of the superfluid stiffness $\rho_s$. The black dashed line denotes the universal $2T/\pi$ line, associated with the BKT transition. **d**, $x$-dependent $T_c$ (red) and $\rho_s$ (blue). $T_c$ is defined as the temperature at which $R_{xx}$ decreases to 50% of its normal-state value at $T = T_{c,onset}$, and the error bars are determined from the difference between $T_{c,onset}$ and $T_{c,0}$.

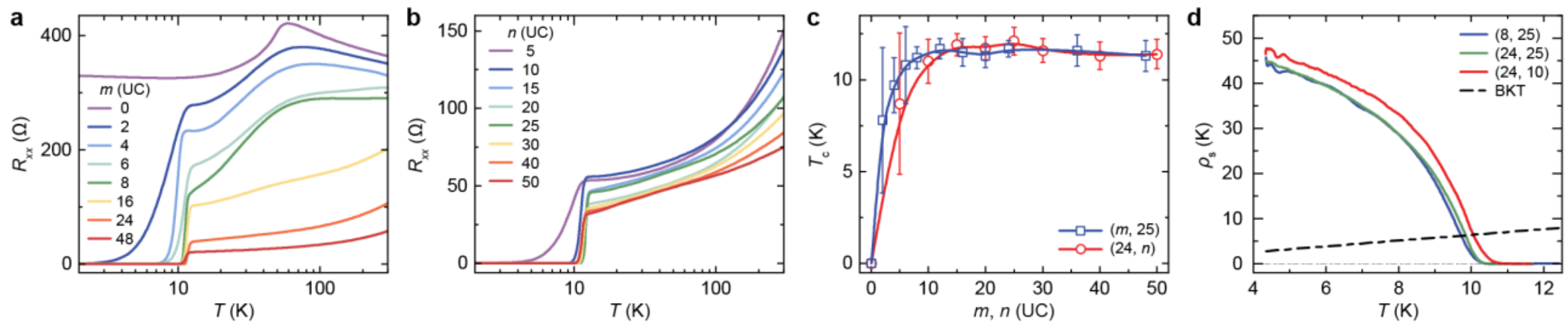

**Fig. 4| Thickness dependence of superconductivity in (*m*, 25) and (24, *n*) bilayers with *x* = 1.** **a**, $R_{xx}$-$T$ curves of the ($m$, 25) bilayers with different $m$. **b**, $R_{xx}$-$T$ curves of the (24, $n$) bilayers with different $n$. **c**, Dependence of $T_c$ on $m$ and $n$ for the ($m$, 25) (blue) and (24, $n$) (red) bilayers, respectively. **d**, $T$ dependence of $\rho_s$ for the (8, 25) (blue), (24, 25) (green), and (24, 10) (red) bilayers. The black dashed line denotes the universal $2T/\pi$ line associated with the BKT transition. The green curve for the (24, 25) bilayer in (**d**) is reused from the red curve ($x$ = 1) in Fig. 3c.